# General System theory, Like-Quantum Semantics and Fuzzy Sets

Ignazio Licata

*Isem, Institute for Scientific Methodology, Pa, Italy*

Ignazio.licata@ejtp.info

**Abstract**: *It is outlined the possibility to extend the quantum formalism in relation to the requirements of the general systems theory. It can be done by using a quantum semantics arising from the deep logical structure of quantum theory. It is so possible taking into account the logical openness relationship between observer and system. We are going to show how considering the truth-values of quantum propositions within the context of the fuzzy sets is here more useful for systemics . In conclusion we propose an example of formal quantum coherence.*

## 1.The role of syntactics and semantics in general system theory

*The omologic element breaks specializations up, forces taking into account different things at the same time, stirs up the interdependent game of the separated sub-totalities, hints at a broader totality whose laws are not the ones of its components. In other words, the omologic method is an anti-separatist and reconstructive one, which thing makes it unpleasant to specialists.*

F. Rossi-Landi 1985

The systemic-cybernetic approach ( Wiener, 1961; von Bertalannfy,1968; Klir, 1991) requires a careful evaluation of epistemology as the critical praxis *internal* to the building up of the scientific discourse. That is why the usual referring to a "connective tissue" shared in common by different subjects could be misleading. As a matter of fact every scientific theory is the outcome of a complex conceptual construction aimed to the problem peculiar features, so what we are interested in is not a framework shaping an abstract super-scheme made by the "filtering" of the particular sciences, but a research focusing on the global and foundational characteristics of scientific activity in a *trans-disciplinary* perspective. According to such view, we can understand the General System Theory (GST) by the analogy to metalogic. It deals with the possibilities and boundaries of various formal systems to a more higher degree than any specific structure.

A scientific theory presupposes a certain set of relations between observer and system, so GST has the purpose to investigate the possibility of describing the multeity of system-observer relationships. The GST main goal is delineating a formal epistemology to study the scientific knowledge formation, a science able to speak about science. Succeeding to outline such panorama will make possible analysing those inter-disciplinary processes which are more and more important in studying complex systems and they will be guaranteed the "transportability" conditions of a modellistic set from a field to another one. For instance, during a theory developing, syntax gets more and more structured by putting univocal constraints on semantics according to the operative requirements of the problem. Sometimes it can be useful generalising a syntactic tool in a new semantic domain so as to formulate new problems. Such work, a typically trans-disciplinary one, can only be done by the tools of a GST able to discuss new relations between syntactics (formal model) and semantics ( model usage). It is here useful to consider again the *omologic perspective*, which not only identifies analogies and isomorphisms in pre-defined structures, but aims to find out a structural and dynamical relation among theories to an higher level of analysis, so providing new use possibilities (Rossi-Landi, 1985). Which thing is particularly useful in studying complex systems, where the very essence of the problem itself makes a dynamic use of models necessary to describe the emergent features of the system (Minati & Brahms, 2002; Collen, 2002).

We want here to briefly discuss such GST acceptation, and then showing the possibility of modifying the semantics of Quantum Mechanics (QM) so to get a conceptual tool fit for the systemic requirements.

## 2. Observer as emergence surveyor and semantic ambiguity solver

*What we look at is not Nature in itself, but Nature unveiling to our questioning methods.*
W. Heisenberg, 1958

A very important and interesting question in system theory can be stated as follows: given a set of measurement systems M and of theories T related to a system S, is it always possible to order them, such that $T_{i-1} \prec T_i$, where the partial order symbol $\prec$ is used to denote the relationship "physically weaker than" ? We shall point out that, in this case, the $i^{th}$ theory of the chain contains more information than the preceding ones. This consequently leads to a second key question: can an unique final theory $T_f$ describe exhaustively each and every aspect of system S ? From the informational and metrical side, this is equivalent to state that all of the information contained in a system S can be extracted, by means of adequate measurement processes.

The fundamental proposition for reductionism is, in fact, the idea that such a theory chain will be sufficient to give a *coherent* and *complete* description for a system S. Reductionism, in the light of our definitions, coincides therefore with the highest degree of semantic space "compression"; each object $D \in T_i$ in S has a definition in a theory $T_i$ belonging to the theory chain, and the latter is - on its turn - related to the fundamental explanatory level of the "final" theory $T_f$. This implies that each aspect in a system S is unambiguously determined by the syntax described in $T_f$. Each system S can be described at a fundamental level, but also with many phenomenological descriptions, each of these descriptions can be considered an approximation of the "final" theory.

Anyway, most of the "interesting" systems we deal with cannot be included in this chained-theory syntax compatibility program: we have to consider this important aspect for a correct epistemic definition of systems "complexity". Let us illustrate this point with a simple reasoning, based upon the concepts of logical openness and intrinsic emergence (Minati, Pessa, Penna, 1998; Licata, 2003b).

Each measurement operation can be theoretically coded on a Turing machine. If a coherent and complete fundamental description $T_f$ exists, then there will also exist a finite set - or, at most, countably infinite - of measurement operations M which can extract each and every single information that describes the system S. We shall call such a measurement set Turing-observer. We can easily imagine Turing-observer as a robot that executes a series of measurements on a system. The robot is guided by a program built upon rules belonging to the theory T. It can be proved, though, that this is only possible for logically closed systems, or at most for systems with a very low degree of logical openness. When dealing with highly logically open systems, no recursive formal criterion exists that can be as selective as requested (i.e., automatically choose which information is relevant to describe and characterize the system, and which one is not), simply because it is not possible to isolate the system from the environment. This implies that the Turing-observer hypothesis does not hold for fundamental reasons, strongly related to Zermelo-Fraenkel's choice axiom and to classical Godel's decision problems. In other words, our robot executes the measurements always following the same syntactics, whereas the scenario showing intrinsic emergence is semantically modified. *So it is impossible thinking to codify any possible measurement in a logically open system!*

The observer therefore plays a key rule, unavoidable as a *semantic ambiguity solver*: only the observer can and will single out intrinsic-observational emergence properties ( Bass & Emmeche,1997; Cariani, 1991), and subsequently plan adequate measurement processes to describe what – as a matter of fact- have turned in  new systems. System complexity is structurally bound to logical openness and is, at the same time, both an expression of highly organized system behaviours (long-range correlations, hierarchical structure, and so on) and an observer's request for new explanatory models.

So, a GST  has to allow - in the very same theoretical context – to deal with the observer as an emergence surveyor in a logical open system. In particular, it is clear that the observer itself is a logical open system.
Moreover, it has to be pointed out that the co-existence of many description levels – compatible but not each other deductible – leads to intrinsic uncertainty situations, linked to the different frameworks  by which a system property can be defined.

## 3. Like-quantum semantics

*I'm not happy with all the analyses that go with just the classical theory, because nature isn't classical, damm it, and if you want to make a simulation of nature, you'd better make it quantum mechanical, and by golly it's a wonderful problem, because it doesn't look so easy. Thank you.*

R. P. Feyman, 1981

When we modify and/or amplify a theory so as to being able to speak about different systems from the ones they were fitted for, it could be better to look at the theory deep structural features so as to get an abstract perspective able to fulfil the omologic approach requirements, aiming to point out a non-banal conceptual convergence.

As everybody knows, the logic of classical physics is a dichotomic language (*tertium non datur*), relatively orthocomplemented and able to fulfil the weak distributivity relations by the logical connectives AND/OR. Such features are the core of the Boolean commutative elements of this logic because disjunctions and conjunctions are symmetrical and associative operations. We shall here dwell on the systemic consequences of these properties. A system S can get or not to get a given property P. Once we fix the P truth-value it is possible to keep on our research over a new preposition P subordinated to the previous one's truth-value. Going ahead, we add a new piece of information to our knowledge about the system. So the relative orthocomplementation axiom grants that we keep on following a successions of steps, each one making our uncertainty about the system to diminish or, in case of a finite amount of steps, to let us defining the state of the system by determining all its properties. Each system's property can be described by a countable infinity of atomic propositions. So, such axiom plays the role of a *describable axiom for classical systems.*

The unconstrained use of such kind of axiom tends to hide the conceptual problems spreading up from the fact that every description implies a context, as we have seen in the case of Turing-observer analysis, and it seems to imply that systemic properties are independent of the observer, it surely is a non-valid statement when we deal with open logical systems. In particular, the Boolean features point out that it is always possible carrying out exhaustively a synchronic description of the properties of a systems. In other words, every question about the system is not depending on the order we ask it and it is liable to a fixed answer we will indicate as 0- false / 1- true. It can be suddenly noticed that the emergent features otherwise get a diachronic nature and can easily make such characteristics not taken for granted. By using Venn diagrams it is possible providing a representation of the complete descriptiveness of a system ruled by classical logics. If the system's state is represented by a point and a property of its by a set of points, then it is always possible a complete "blanketing" of the universal set I, which means the always universally true proposition. (see fig. 1).

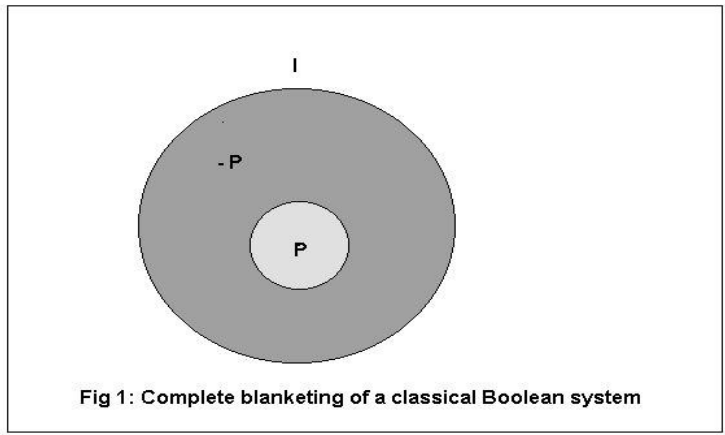

Fig 1: Complete blanketing of a classical Boolean system

The quantum logics shows deep differences which could be extremely useful for our goals (Birkhoff & von Neumann, 1936; Piron, 1964). At the beginning it was born to clarify some QM's counter-intuitive sides, later it has developed as an autonomous field greatly independent from the matters which gave birth to it. We will abridge here the formal references to an essential survey, focusing on some points of general interest in systemics.

The quantum language is a non-Boolean orthomodular structure, which is to say it is relatively orthocomplemented but *non-commutative*, for the crack down of the distributivity axiom. Such thing comes naturally from the *Heisenberg Indetermination Principle* and binds the truth-value of an assertion to the context and the order by which it has been investigated (Griffiths, 1995). A well-known example is the one of a particle's spin measurement along a given direction. In this case we deal with semantically well defined possibilities and yet intrinsically uncertain. Let put $\Psi x$ the spin measurement along the direction x. For the indetermination principle the value $\Psi y$ will be totally uncertain, yet the proposition $\Psi y = 0 \lor \Psi y = 1$ is necessarily true. In general, if P is a proposition, (-P) its negation and Q the property which does not commute with P, then we will get a situation that can be represented by a "patchy" blanketing of the set I (see fig.2).

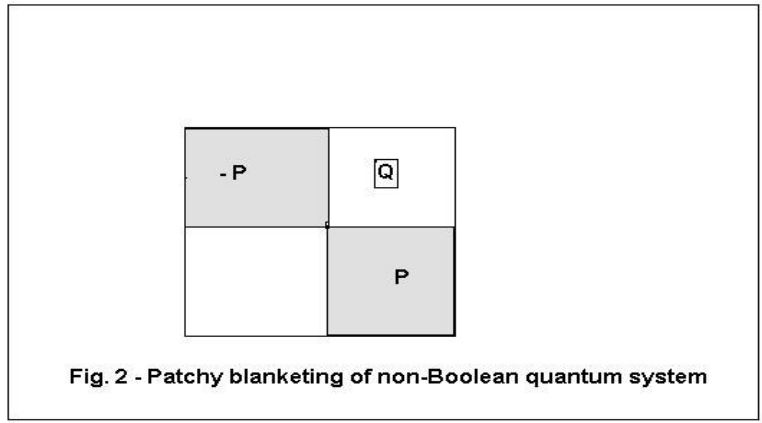

Fig. 2 - Patchy blanketing of non-Boolean quantum system

Such configuration finds its essential meaning just in its relation with the observer. So we can state that when a situation can be described by a quantum logics, a system is never completely defined a priori. The measurement process by which the observer's action takes place is a choice fixing some system's characteristics and letting other ones undefined. It happens just for *the nature itself of the observer-system inter-relationship.* Each observation act gives birth to new descriptive possibilities. The proposition Q – in the above example – describes properties that cannot be defined by any implicational chain of propositions P. Since the intrinsic emergence cannot be regarded as a system property independent of the observer action- as in naïve classical emergentism - , Q can be formally considered the expression of an emergent property. *Now we are strongly tempted to define as emergent the undefined proposition of quantum-like anti-commutative language.* In particular, it can be showed that a non-Boolean and irreducible orthomodular language arises infinite propositions. It means that for each couple of propositions $P_1$ and $P_2$ such that non of them imply the other, there exists infinite propositions Q which imply $P_1 \vee P_2$ without necessarily implying the two of them separately: *tertium datur*. In a sense, the disjunction of the two propositions gets more information than their mere set-sum, that is the entirely opposite of what happens in the Boolean case. It is now easy to comprehend the deep relation binding the anti-commutativity, indetermination principles and system's holistic global structure. A system describable by a Boolean structure can be completely "solved" by analysing the sub-systems defined by a fit decomposition process( Heylighen, 1990; Abram, 2002). On the contrary, in the anti-commutative case studying any sub-system modifies the entire system in an irreversible and structural way and produces uncertainty correlated to the gained information, which think makes absolutely natural extending the indetermination principles to a big deal of spheres of strong interest for systemics (Volkenshtein , 1988).

A particularly key-matter is how to conceptually managing the infinite cardinality of emergent propositions in a lik-quantum semanics. As everybody knows traditional QM refers to the frequentistic probability worked out within the Copenhagen Interpretation (CIQM). It is essentially a *sub specie probabilitatis* Boolean logics extension. The values between $[0,1]$ - i.e. between the completely and always true proposition I and the always false one O – are meant as expectation values, or the probabilities associated to any measurable property. Without dwelling on the complex – and as for many questions still open – debate on QM interpretation, we can here ask if the probabilistic acception of truth-values is the fittest for system theory. As it usually happens when we deal with trans-disciplinary feels, it will bring us to add a new, and of remarkable interest for the "ordinary" QM too, step to our search.

**4. A Fuzzy Interpretation of Quantum Languages**

*A slight variation in the founding axioms of a theory can give way to huge changings on the frontier.*
S. Gudder, 1988

The study of the structural and logical facets of quantum semanics does not provide any necessary indications about the most suitable algebraic space to implement its own ideas. One of the thing which made a big merit of such researches has been to put under discussion the key role of Hilbert space. In our approach we have kept the QM "internal" problems and its extension to systemic questions well separated. Anyway, the last ones suggest an interpretative possibility bounded to fuzzy logic, which thing can considerably affect the traditional QM too. The *fuzzy set theory* is , in its essence, a formal tool created to deal with information characterized with vagueness and indeterminacy. The by-now classical paper of Lotfi Zadeh (Zadeh, 1965) brings to a conclusion an old tradition of logics, which counts Charles S. Peirce, Jan C. Smuts, Bertrand Russell, Max Black and Ian Lukasiewicz among its forerunners. At the core of the fuzzy theory lies the idea that an element can belong to a set to a variable degree of membership; the same goes for a proposition and its variable relation to the true and false logical constants. We underline here two aspects of particular interest for our aims. The fuzziness' definition concerns single elements and properties, but not a statistical ensemble, so it has to be considered a completely different concept from the probability one, it should –by now- be widely clarified (Mamdani, 1977; Kosko, 1990). A further essential – even maybe less evident – point is that fuzzy theory calls up a non-algorithmic "oracle", an observator (i.e. a logical open system and a semantic ambiguity solver) to make a choice as for the membership degree. In fact, the most part of the theory in its structure is free-model; no equation and no numerical value create constraints to the quantitative evaluation, being the last one the model builder's task. There consequently exists a deep bound between systemics and fuzziness successfully expressed by the *Zadeh's incompatibility principle* (Zadeh, 1972) which satisfies our requirement for a generalized indeterminacy principle. It states that by increasing the system complexity (i.e. its logical openness degree), it will decrease our ability to make exact statements and proved predictions about its behaviour. There already exists many examples of crossing between fuzzy theory and QM (Dalla Chiara, Giuntini, 1995; Cattaneo, Dalla Chiara, Giuntini 1993). We want here to delineate the utility of fuzzy polyvalence for systemic interpretation of quantum semantics.

Let us consider a complex system, such as a social group, a mind and a biological organism. Each of these cases show typical emergent features owed both to the interaction among its components and the inter-relations with the environment. An act of the observer will fix some properties and will let some others undetermined according to a non-Boolean logic. The recording of such properties will depend on the succession of the measurement acts and their very nature. The kind of complexity into play, on the other hand, prevents us by stating what the system state is so as to associate to the measurement of a property an expectation probabilistic value. In fact, just the above-mentioned examples are related to macroscopic systems for which the probabilistic interpretation of QM is patently not valid. Moreover, the traditional application of the probability concept implies the notion of "possible cases", and so it also implies a pre-defined knowledge of systems' properties. However, the non-commutative logical structure here outlined does not provide any cogent indication on probability usage.

Therefore, it would be proper to look at a fuzzy approach so to describe the measurement acts. We can state that *given a generic system endowed with high logical openness and an indefinite set of properties able of describing it, each of them will belong to the system in a variable degree.* Such viewpoint expressing the famous theorem of fuzzy "subsetness" – also known as "the whole into the part" principle – could seem to be too strong , indeed it is nothing else than the most natural expression of the actual scientific praxis facing intrinsic emergent systems. At the beginning, we have at our disposal indefinite information progressively structuring thanks to the feedback between models and measurements. It can be shown that any logically open model of degree n – where n is an integer – will let a wide range of properties and propositions indeterminate (the Qs in fig. 2).The above-mentioned model is a "static" approximation of a process showing aspects of variable closeness and openness. The latter ones varies in time, intensity, different levels and context. It is remarkable pointing out how such systems are "flexible" and context-sensitive, change the rules and make use of "contradictions" . This point has to be stressed to understand the link between fuzzy logic and quantum languages. By increasing the logical openness and the unsharp properties of a system, it will be less and less fit to be described by a Boolean logic. It brings as a consequence that for a complex system the intersection between a set (properties, propositions) and its complement is not equal to the empty set, but it includes they both in a fuzzy sense. So we get a polyvalent semantic situation which is well fitted for being described by a quantum language. As for our systemic goal it is the probabilistic interpretation to be useless, so we are going to build a fuzzy acception of the semantics of the formalism. In our case, *given a system S and a property Q,, let $\Psi$ be a function which associates Q to S, the expression $\Psi_S(Q) \in [0,1]$ has not to be meant as a probability value, but as a degree of membership.* Such union between the non-commutative sides of quantum languages and fuzzy polyvalence appears to be the most suitable and fecund for systemics.

Let us consider the traditional expression of quantum coherence (the property expressing the QM global and non-local characteristics, i.e. superposition principle, uncertainty, interference of probabilities), $\Psi = a_1\Psi_1 + a_2\Psi_2$. In the fuzzy interpretation, it means that the properties $\Psi_1$ e $\Psi_2$ belong to $\Psi$ with degrees of membership $a_1$ e $a_2$ respectively. In other words, *for complex systems the Schrödinger's cat can be simultaneously both alive and dead !* Indeed the recent experiments with SQUIDs and the other ones investigating the so-called macroscopic quantum states suggest a form of macro-realism quite close to our fuzzy acception (Leggett, 1980; Chiatti, Cini, Serva, 1995). It can provide *in nuce* an hint which could show up to be interesting for the QM old-questioned interpretative problems.

In general, let x be a position coordinate of a quantum object and $\Psi$ its wave function, $|\Psi(x)|^2 dV$ is usually meant as the probability of finding the particle in a region dV of space. On the contrary, in the fuzzy interpretation we will be compelled to look at the $\Psi$ square modulus as the degree of membership of the particle to the region dV of space. How unusual it may seem, such idea has not to be regarded thoughtlessly at. As a matter of fact, in Quantum Field Theory and in other more advanced quantum scenarios, a particle is not only a localized object in the space, but rather an event emerging from the non-local networks elementary quantum transition (Licata, 2003a). Thus, the measurement is a "defuzzification" process which, according to the stated, reduces the system ambiguity by limiting the semantic space and by defining a fixed information quantity.

If we agree with such interpretation we will easily and immediately realize that we will able to observe quantum coherence behaviours in non-quantum and quite far from the range of Plank's h constant situations. We reconsider here a situation owed to Yuri Orlov (Orlov, 1997).

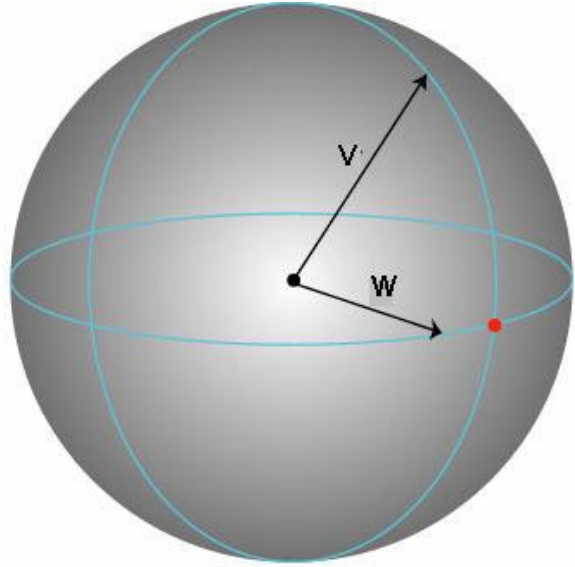

Fig.3 A Riemann' sphere built on an Argand's plane. V and W represent complex amplitudes

Let us consider a Riemann's sphere (Dirac, 1947) – see fig. 3 - and let assume that each point on the sphere represents a single interpretation of a given situation, i.e. the assigning of a coherent set of truth-values to a given proposition. Alternatively, we can consider the choosing of a vector $\mathbf{v}$ from the centre O to a point on the sphere as a logical definition of a world. If we choose a different direction, associated to a different vector $\mathbf{w}$, we can now set the problem about the meaning of the amplitude between the logical descriptions of the two worlds. It is known that such amplitude is expressed by $\frac{1}{2}(1+\cos\vartheta)$, where $\vartheta$ is the angle between the two interpretations. The amplitude corresponds to a superposition of worlds, so producing the typical interference patterns which in vectorial terms are related to $\mathbf{w}/\mathbf{v}$. In this case, the traditional use of probability is not necessary because our knowledge of one of the two world with probability equal to p =1 (certainity), say nothing us about the other one probability. An interpretation is not a quantum object in the proper sense, and yet we are forced to formally introduce a wave-function and interference terms whose role is very obscure a one. The fuzzy approach, instead, clarifies the quantum semantics of this situation by interpreting interference as a measurement where the properties of the world $v|\Psi_v\rangle + w|\Psi_w\rangle$ are owed to the global and indissoluble (non-local) contribution of the $\mathbf{v}$ and $\mathbf{w}$ overlapping.

In conclusion, the generalized using of quantum semantics associated to new interpretative possibilities gives to systemics a very powerful tool to describe the observator-environment relation and to convey the several, partial attempts - till now undertaken - of applying the quantum formalism to the study of complex systems into a comprehensive conceptual root.

**ACKNOWLEDGEMENTS**

A special thank to Prof. G. Minati for his kindness and his supporting during this paper drafting. I owe a lot to the useful discussing on structural Quantum Mechanics and logics with my good friends Prof. Renato Nobili (who let me use the figs. 1 and 2 from his book "Dai Quark alla Mente", to be published) and Prof. Eliano Pessa. Dedicated to M.V.